\documentclass[lettersize,journal]{IEEEtran}
\usepackage{amsmath,amsfonts}
\usepackage{algorithmic}
\usepackage{algorithm}
\usepackage{amsmath, amssymb}
\usepackage{pifont}
\usepackage{array}
\usepackage[caption=false,font=normalsize,labelfont=sf,textfont=sf]{subfig}
\usepackage{textcomp}
\usepackage{stfloats}
\usepackage{url}
\usepackage{verbatim}
\usepackage{graphicx}
\usepackage{cite}
\usepackage{hyperref}
\usepackage{amssymb}
\usepackage{mathrsfs}
\usepackage{soul}
\usepackage{color, xcolor}
\begin{document}

\title{Active RIS Enabled NLoS LEO Satellite Communications:\\A Three-timescale Optimization Framework}

\author{Ziwei Liu,~\IEEEmembership{Senior Member,~IEEE,} Junyan He, Shanshan Zhao,~\IEEEmembership{Member,~IEEE}, Meng Hua,~\IEEEmembership{Member,~IEEE},\\ Bin Lyu,~\IEEEmembership{Senior Member,~IEEE,} Xinjie Zhao, \IEEEmembership{Student Member,~IEEE}, and Gengxin Zhang, \IEEEmembership{Member,~IEEE}
\thanks{Ziwei Liu, Junyan He,  Shanshan Zhao, Bin Lyu, Xinjie Zhao and Gengxin Zhang, are with Nanjing University of Posts and Telecommunications, Nanjing 210003, China (e-mail:\{lzw,1023010438\}@njupt.edu.cn; zhaoshanshan025@163.com; \{blyu,2022010108,zgx\}@njupt.edu.cn). Meng Hua is with the Department
	of Electrical and Electronic Engineering, Imperial College London,
	London SW7 2AZ, UK (e-mail: m.hua@imperial.ac.uk).}
\thanks{}
}
\markboth{}%
{Shell \MakeLowercase{\textit{et al.}}: A Sample Article Using IEEEtran.cls for IEEE Journals}

\maketitle

\begin{abstract}
In this letter, we study an active reconfigurable intelligent surfaces (RIS) assisted Low Earth orbit (LEO) satellite communications under non-line-of-sight (NLoS) scenarios, where the active RIS is deployed to create visual line-of-sight links for reliable communication. To address the challenges of high energy consumption caused by frequent beamforming updates in active RIS, we propose a three-timescale optimization framework that jointly designs the transmit beamforming, RIS beamforming, and RIS direction vectors based on their characteristics. The goal is to maximize the system achievable rate while reducing energy consumption by controlling the RIS beamforming switching frequency. Then, a two-layer solution framework is developed, incorporating fractional programming (FP), alternating optimization (AO), successive approximation (SCA), and penalty-based methods, to obtain the optimized solution. Simulation results demonstrate that the proposed scheme can effectively improve system performance and reduce the energy consumption of the active RIS.

\end{abstract}


\section{Introduction}
\IEEEPARstart{R}{ecently}, Low Earth Orbit (LEO) satellite systems, owing to their wide communication coverage and low transmission latency, have been considered as a critical supplement to terrestrial networks in future 6G communication systems \cite{ref1}. However, the quality of satellite communication services can degrade owing to link blockages, which hinders reliable communication in environments such as canyons and complex urban areas where only non-line-of-sight (NLoS) channels are available. 

The recent emergence of the reconfigurable intelligent surfaces (RIS), which can intelligently control electromagnetic wave propagation to create virtual line-of-sight (VLoS) links, provides a novel solution due to their low energy consumption and ease of deployment\cite{ref2}. A continuous-time propagation model for RIS-assisted LEO satellite communication was proposed in \cite{ref3}, demonstrating that introducing a RIS-assisted link can enhance system performance and reliability. In \cite{ref4}, the authors employed RIS to establish VLoS links in urban NLoS environments and qualitatively showed the importance of RIS deployment direction. The authors in \cite{ref5} investigated the maximum system sum rate in RIS-aided multi-user ultra-dense LEO satellite networks. However, prior studies typically assumed passive RIS deployed on fixed infrastructure, which requires a large number of RIS elements to combat the double fading effect \cite{ref6}. In this regard, active RIS with fewer elements has been proposed to enhance the system performance in LEO satellite communications through signal amplification \cite{ref7}. However, most existing works on RIS, i.e., \cite{ref3,ref4,ref5,ref6,ref7}, neglected the impact of optimizing RIS deployment direction on system performance. Moreover, the RIS beamforming (also known as RIS phase shifts) is typically designed based on the instantaneous channel state information (CSI)\cite{ref8}. This increases the energy consumption for switching RIS phase shifts, making it difficult to meet the long-term communication demands in energy-constrained scenarios such as canyons and disaster-affected areas. Thus, it is crucial to develop  multi-timescale optimization strategies that account for the inherent characteristics of each parameter within RIS-assisted satellite systems.

Motivated by the above, this letter proposes a three-timescale optimization framework for active RIS-aided LEO satellite communication systems under NLoS conditions. In this framework, the RIS deployment direction is updated at the beginning of the communication period, which consists of multiple holding intervals. During each holding interval, the RIS beamforming vector is optimized across several unit time slots, while the satellite transmit beamforming is updated in every time slot. To this end, we formulate an achievable rate maximization problem and propose a two-layer solution framework to find a sub-optimal solution. In the inner-layer, we adopt a hybrid algorithm combining the fractional programming (FP) framework \cite{ref9}, alternating optimization (AO) algorithm \cite{ref10}, successive approximation
(SCA) \cite{ref11}, and the penalty function methods \cite{ref12} to jointly optimize above variables; In the outer-layer, we adjust the RIS beamforming switching frequency (i.e., the number of holding intervals) to reduce the active RIS energy consumption. Numerical results show the effectiveness of the proposed scheme and reveal how the active RIS energy budget affects overall system performance.
\section{SYSTEM MODEL}
\begin{figure}[!t]
	\centering
	\includegraphics[width=2.5in]{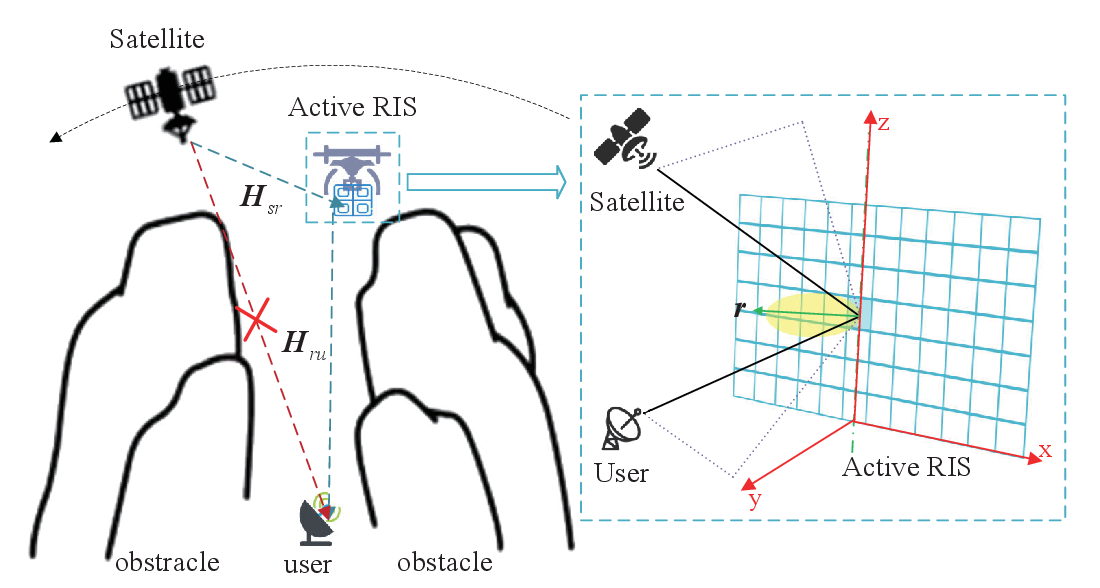}
	\caption{Active RIS-aided LEO Satellite communication systems.}
	\label{fig_1}
	\vspace{-4mm} 
\end{figure}

As shown in Fig. \ref{fig_1}, we consider a single-input single-output (SISO) downlink LEO satellite communication system in a canyon scenario, where obstacles (e.g., mountains) block the LoS link between the satellite and the user. To ensure reliable communication from the satellite with $L$ antennas to the user with a single-antenna, we adopt an active RIS mounted on the unmanned aerial vehicle and deployed in the vicinity of the user to build a VLoS link. This aerial deployment is particularly practical for the canyon environment, where traditional infrastructure is difficult to build \cite{ref13}. Due to the reflective nature of the active RIS, the satellite and the user are located on the same side of the active RIS to ensure valid signal reflection. Without loss of generality, we assume that the user remains relatively static, while the LEO satellite exhibits high mobility\cite{ref3}.
\subsection{Active RIS Modeling}
The active RIS comprises $M$ rows and $N$ columns of reflective elements with a size of $\lambda/2$, where $\lambda$ denotes the carrier wavelength. The active RIS reflection coefficient matrix is $\mathbf{\Theta}= \text{diag}(A_{1,1}e^{j\boldsymbol{\phi}_{1,1}},\dots, A_{M,1}e^{j\boldsymbol{\phi}_{M,1}},\dots, A_{M,N}e^{j\boldsymbol{\phi}_{M,N}})$, where ${{\boldsymbol{A}_{m,n}}\in[0,a_{max}]}$ and ${\boldsymbol{\phi}_{m,n}\in[0,2\pi]}$ denote the amplification coefficient and phase shift of the reflective element located in the $m$-th row and $n$-th column, respectively. The parameter ${a_{max}\geq1}$ denotes the maximum amplitude factor. 

According to \cite{ref1} and \cite{ref2}, let ${{F}_{RIS}}(\boldsymbol{r},\boldsymbol{l})$ denote the normalized power radiation pattern of the active RIS, which can be expressed as
\begin{equation}
	\label{Formulas1}
	{{F}_{RIS}}(\boldsymbol{r},\boldsymbol{l})=\left\{ \begin{matrix}
		{(\boldsymbol{r}^T\boldsymbol{l})}^{\beta}  & \quad 0\le\boldsymbol{r}^T\boldsymbol{l}\le1,  \\
		\quad0 & -1\le\boldsymbol{r}^T\boldsymbol{l} < 0,  \\
	\end{matrix} \right.
\end{equation}
where ${\boldsymbol{r}\in\mathbb{R}^{3\times1}}$ and ${\boldsymbol{l}\in\mathbb{R}^{3\times1}}$ represent unit vectors of the RIS deployment direction and signal propagation path, respectively. The exponent $\beta$ denotes the radiation exponent and is set to 1 for simplicity.
\subsection{Signal Transmission Modeling}
\begin{figure}[!t]
	\centering
	\includegraphics[width=2.5in]{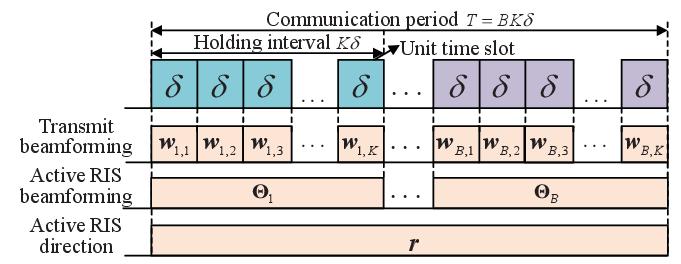}
	\caption{Illustration of three-timescale framework.}
	\label{fig_2}
	\vspace{-6mm} 
\end{figure}

As illustrated in Fig. \ref{fig_2}, we consider a three-timescale framework to model the signal transmission, aiming to balance the system performance and design complexity. Specifically, the communication period with duration $T$ is divided into $B$ holding intervals, each consisting of $K$ unit time slots of $\delta=1 \text{s}$. Here, the values of $K$ and $B$ can be dynamically adjusted to maintain the total communication period duration of $T=BK\delta$ unchanged. Within each unit time slot, the satellite position and channel state are assumed to remain constant\cite{ref5}. Given the real-time adjustability of the satellite precoders, we assume that the transmit beamforming at the satellite, denoted by $\{{\boldsymbol{w}_{b,k}\}_{b\in \mathcal{B}, k\in\mathcal{K}}\in\mathbb{C}^{L\times 1}}$, updates at every unit time slot, where $\mathcal{B}\in\{1,...,B\}$ and $\mathcal{K}\in\{1,...,K\}$ are the sets of holding intervals within a communication period and unit time slots per holding interval, respectively. Due to the difficulty of adjusting the RIS orientation after UAV deployment, the RIS direction vector $\boldsymbol{r}$ remains constant throughout the communication period. For the active RIS beamforming matrix $\{\mathbf{\Theta}_b\}_{b\in\mathcal{B}}$, we update it at every holding interval and adjust the holding interval duration to enable control over the beamforming switching frequency, thereby reducing the active RIS switching and control power consumption \cite{ref6}.

Let ${{\boldsymbol{H}}_{sr}\in {{\mathbb{C}}^{NM\times L}}}$ and ${{\boldsymbol{H}}_{ru}\in {{\mathbb{C}}^{NM\times 1}}}$ denote the channels from the satellite to the active RIS (i.e., Sat-aRIS channel), and from the active RIS to the user (i.e., aRIS-User channel), respectively. Considering the atmospheric effects and free space path loss, the Sat-aRIS channel is modeled as ${\boldsymbol{H}}_{sr}= \sqrt{C_{sr}}\xi^{- \frac{1}{2}}\boldsymbol{b}^{\frac{1}{2}} \odot e^{j\boldsymbol{\psi}}$, where ${C_{sr}}=(\frac{\lambda}{4\pi{d_{sr}}})^2$ with ${d}_{sr}$ being the distance between the satellite and the active RIS, $\boldsymbol{b}$ denotes the beam gain vector of the satellite as given in \cite{ref14}, $\boldsymbol{\psi}$ denotes the random phase matrix over $[0,2\pi)$, $\odot$ is the Hadamard product, $\xi$ represents the rain attenuation coefficient and satisfies ${\ln(\xi_{dB})\sim\mathcal{CN}(\mu,\sigma^2)}$. The Doppler effect caused by LEO satellite mobility is assumed to be compensated perfectly \cite{ref1}. The aRIS-User channel is modeled as the Rician channel to account for the multipath effect, which is given by ${\boldsymbol{H}_{ru}}=\sqrt{\frac{{{\kappa }}}{{{\kappa }}+1}}{{C}_{ru}}+\sqrt{\frac{1}{{{\kappa }}+1}}{{C}_{ru}}e^{j\boldsymbol{\psi}}$, where $\kappa$ denotes the Rician factor, ${C}_{ru}=(\frac{\lambda}{4\pi{d_{ru}}})^2$  is free space loss.

Following \cite{ref2} and \cite{ref6}, the received signal at the user in the $k$-th unit time slot of the $b$-th holding interval is given by $y_{b,k}= c_1\sqrt{\boldsymbol{r}^{\text{T}}\boldsymbol{l}_{sr}\boldsymbol{r}^{\text{T}}\boldsymbol{l}_{ru}}\boldsymbol{{H}}_{ru}^{\text{H}}\boldsymbol{\Theta}_{b}{\boldsymbol{H}_{sr}}\boldsymbol{w}_{b,k}s_{b,k}+c_2\sqrt{\boldsymbol{r}^{\text{T}}\boldsymbol{l}_{ru}}\boldsymbol{H}_{ru}^{\text{H}}\boldsymbol{\Theta}_b\boldsymbol{n}_{1}+{n}_{2}$, where $\boldsymbol{l}_{sr}$ and $\boldsymbol{l}_{ru}$ respectively denote the unit vectors of the Sat-aRIS and aRIS-User signal propagation path as defined in (\ref{Formulas1}), ${\boldsymbol{n}_1\in {{\mathbb{C}}^{NM\times 1}}}$ denotes the thermal noise at the active RIS, each element of which follows $\mathcal{CN}(0,\sigma_1^2)$, ${n_2\sim\mathcal{CN}(0,\sigma_2^2)}$ is the Gaussian noise at the user, and ${s_{b,k}\sim \mathcal{CN}(0,1)}$ is the satellite transmit signal. The constants ${c}_{1}$ and ${c}_{2}$ are defined as ${{c}_{1}=\sqrt{\pi{G}_{R}{{G}_{U}}{{G}_{S}}{{P}_{S}}s_{b,k}}}$, ${{c}_{2}=\sqrt{\frac{4\pi }{{{\lambda }^{2}}}{G}_{R}{{G}_{U}}}}$, where  ${{G}_{U}}$ and ${{G}_{S}}$ denote the antenna gains of the user and the satellite, respectively, ${G}_{R}$ is the antenna gain of the active RIS as given in \cite{ref2}, and ${{P}_{S}}$ represents the satellite's transmit power. The received  signal-to-noise ratio at the user can be formulated as $\gamma_{b,k}=\frac{|{{c}_{1}}\sqrt{{{\boldsymbol{r}}^{\text{T}}}{\boldsymbol{{l}}_{sr}}{\boldsymbol{r}^{\text{T}}}\boldsymbol{{l}}_{ru}}{{\boldsymbol{H}{{_{ru}^\text{H}}\mathbf{\Theta}}_{b}}}\boldsymbol{H}_{sr}\boldsymbol{w}_{b,k}|^2}{{{c}_{2}^{2}{\boldsymbol{r}^{\text{T}}}\boldsymbol{{l}}_{ru}\left\|{{\boldsymbol{H}_{ru}^\text{H}\boldsymbol{\Theta}}_{b}}\right\|^{2}}\sigma _{1}^{2}+\sigma_{2}^{2}}$. The instantaneous achievable rate in the $k$-th unit time slot of the $b$-th holding interval is given by $R_{b,k}=\log_2 (1+\gamma_{b,k})$. Accordingly, the system average achievable rate through the communication period is given by $\overline{R}={\frac{1}{BK}}\sum\nolimits_{b=1}^{B}\sum\nolimits_{k=1}^{K}R_{b,k}$. According to \cite{ref6}, the energy consumption of the active RIS over the communication period is modeled as $E=\sum\nolimits_{b=1}^{B}\sum\nolimits_{k=1}^{K}\,[{{\eta}}\left\|{c_{1}}\sqrt{{\boldsymbol{r}^{\text{T}}}{\boldsymbol{l}_{sr}}}{{{\boldsymbol{\Theta }}_{b}}\boldsymbol{H}_{sr}{\boldsymbol{w}_{b,k}}}\right\|^{2}+{{\eta}}\left\|{\boldsymbol{\Theta}_{b}}\right\|^{2}\sigma_{1}^{2}]+BMN{{P}_{C}}$, where $\eta$ is the reciprocal of amplifier efficiency, ${{P}_{\text{C}}}$ represents the switching and control power consumption of the active RIS (incurred during beamforming switching). As discussed in \cite{ref6}, the DC bias loss of the active RIS remains constant over the communication period and is therefore omitted from the energy consumption analysis.
\section{Problem Formulation and Solution}
\subsection{Problem Formulation}
In this section, we investigate and formulate the joint optimization of the satellite transmit beamforming, the beamforming and direction vector of the active RIS, and the holding interval duration to maximize the system achievable rate defined as follows:
\begin{equation}
	\begin{aligned}
		\label{Formulas2}
		&\max_{\{{\boldsymbol{w}_{b,k}\}_{b\in\mathcal{B}, k\in\mathcal{K}}},{\{\mathbf{\Theta}_b\}_{b\in\mathcal{B}}},\boldsymbol{r},K} \quad\overline{R}  \\
		&\quad\quad\quad\quad\quad\text{s.t.} 
		\begin{array}[t]{l}
			\text{C1:} \quad {\left\|\boldsymbol{r}\right\|}^{2}=1, \\
			\text{C2:} \quad 0\le {\boldsymbol{r}^{\text{T}}}{\boldsymbol{l}_{sr}}\le 1, 0\le {\boldsymbol{r}^{\text{T}}}{\boldsymbol{l}_{ru}}\le 1,\\
			\text{C3:}  \quad {\left\|\boldsymbol{w}_{b,k}\right\|}^{2}=1,\; \forall b, \;\forall k, \\
			\text{C4:}  \quad E\le {E_{max}}, \\
			\text{C5:}  \quad 0\le{\boldsymbol{A}_{m,n}}\le {{a}_{max}}. \\
		\end{array}
	\end{aligned}
\end{equation} 

In problem (\ref{Formulas2}), constraint $\text{C1}$ ensures the direction vector of the active RIS is a unit normal vector. Constraint $\text{C2}$ guarantees the validity of the signal's incident and reflection angles. Constraint $\text{C3}$ denotes the normalization constraint for the transmit beamforming at the satellite. $\text{C4}$ represents the constraint of energy consumption, where $E_{max}$ is the maximum energy budget. Constraint $\text{C5}$ is the amplitude constraint. To tackle the challenges arising from the non-convexity of the objective function and variable coupling, a two-layer solution framework is proposed for solving problem ({\ref{Formulas2}}). In the outer-layer, we utilize an iterative method to determine the optimal holding interval duration under constraint C4. In the inner-layer, we decompose the original problem into three subproblems and propose an AO algorithm with the FP technique to iteratively optimize ${{\{\boldsymbol{w}_{b,k}\}}_{b\in\mathbf{B}, k\in\mathbf{K}}}$, ${\{\mathbf{\Theta}_b\}_{b\in\mathbf{B}}}$ and  $\boldsymbol{r}$.
\subsection{Inner-layer Optimization}\label{sec:III-B}
Given a fixed value of $K$, we first introduce the inner-layer optimization, which is composed of the optimization of ${\{\boldsymbol{w}_{b,k}\}_{b\in\mathcal{B}, k\in\mathcal{K}}}$, $\{\mathbf{\Theta}_b\}_{b\in\mathcal{B}}$, and $\boldsymbol{r}$.
\subsubsection{Optimization of Transmit Beamforming ${\{\boldsymbol{w}_{b,k}\}_{b\in\mathcal{B}, k\in\mathcal{K}}}$}
With $\{\mathbf{\Theta}_b\}_{b\in\mathcal{B}}$ and $\boldsymbol{r}$ fixed, we first optimize the transmit beamforming at the satellite. Since all $\{\boldsymbol{w}_{b,k}\}_{b\in\mathcal{B}, k\in\mathcal{K}}$ in different unit time slots are independent,
the subproblem reduces to optimizing $\boldsymbol{w}_{b,k}$ individually for each $b\in\mathcal{B}$ and $k\in\mathcal{K}$, given by
\begin{equation}
		\label{Formulas3}
		\max_{\boldsymbol{w}_{b,k}} \quad {c}_{3}{\boldsymbol{w}^\text{H}_{b,k}}{\boldsymbol{H}_1}{\boldsymbol{w}_{b,k}},\quad\text{s.t.}\;\text{C3},
\end{equation}
where $\boldsymbol{H}_1=(\boldsymbol{H}_{sr}^{\text{H}}{\boldsymbol{\Theta }}_{b}\boldsymbol{H}_{ru})(\boldsymbol{H}_{ru}^{\text{H}}{{\boldsymbol{\Theta}}_{b}}\boldsymbol{H}_{sr})$ and ${c}_{3}={c}_{1}^2{\boldsymbol{r}^{\text{T}}}{\boldsymbol{l}_{sr}}{\boldsymbol{r}^{\text{T}}}{\boldsymbol{l}_{ru}}$. By using the Rayleigh quotient\cite{ref15}, the optimal solution to (\ref{Formulas3}) is constructed as ${{\boldsymbol{w}}_{b,k}^{*}}=\frac{\boldsymbol{\ell} }{||\boldsymbol{\ell} ||}$,
where $\boldsymbol{\ell}$ represents the eigenvector corresponding to the maximum eigenvalue of $\boldsymbol{H}_1$.

\subsubsection{Optimization of active RIS Beamforming $\{\mathbf{\Theta}_b\}_{b\in\mathcal{B}}$}\label{sec:III-B2}
With the obtained ${\{\boldsymbol{w}_{b,k}\}_{b\in\mathcal{B}, k\in\mathcal{K}}}$ and fixed $\boldsymbol{r}$, we proceed to optimize the active RIS beamforming using the FP framework \cite{ref9}. Specifically, we introduce auxiliary variables ${\boldsymbol{v}_{b,k}}\in{\mathbb{C}}^{BK\times 1}$ and ${\boldsymbol{\mu}_{b,k}}\in{\mathbb{C}}^{BK\times 1}$, and employ both the quadratic transform and the Lagrangian dual transform to decouple the logarithmic terms in the multi-ratio objective function of (\ref{Formulas2}). Accordingly, the corresponding subproblem can be formulated as
\begin{equation}
	\begin{aligned}
		\label{Formulas4}
		\max_{{{\mathbf{\Theta }}_{b}},{\boldsymbol{v}_{b,k}},{\boldsymbol{\mu}_{b,k}}}&f({{\mathbf{\Theta }}_{b}},{\boldsymbol{v}_{b,k}},{\boldsymbol{\mu }_{b,k}})=\sum\limits_{b=1}^{B}{\sum\limits_{k=1}^{K}{{{\log }_{2}}(1+{\boldsymbol{v}_{b,k}})}}-\\
		&{\boldsymbol{v}_{b,k}}+2{\boldsymbol{\mu }_{b,k}}\sqrt{1+{\boldsymbol{v}_{b,k}}}{{c}_{4}}\boldsymbol{H}_{ru}^{\text{H}}{{\boldsymbol{\Theta }}_{b}}{\boldsymbol{H}_{sr}}{\boldsymbol{w}_{b,k}}-\mu _{b,k}^{2}\\
		&(\boldsymbol|{{c}_{4}}\boldsymbol{H}_{ru}^{\text{H}}{{\boldsymbol{\Theta }}_{b}}{\boldsymbol{H}_{sr}}{\boldsymbol{w}_{b,k}}{{|}^{2}}+c_{5}^2{\left\|{\boldsymbol{H}_{ru}^\text{H}}{{\boldsymbol{\Theta }}_{b}}\right\|}^2\sigma_1^{2}+\sigma_2^{2})\\
		\text{s.t.}\quad& \text{C4} , \text{C5},
	\end{aligned}
\end{equation} 
where ${{c}_{4}}={{c}_{1}}\sqrt{{\boldsymbol{r}^{\text{T}}}{\boldsymbol{l}_{sr}}{\boldsymbol{r}^{\text{T}}}{\boldsymbol{l}_{ru}}}$, ${{c}_{5}}={{c}_{2}} \sqrt{{\boldsymbol{r}^{\text{T}}}{\boldsymbol{l}_{ru}}}\sigma _{1}$. We iteratively optimize the variables in (\ref{Formulas4}). Specifically,
when $\mathbf{\Theta }_{b}$ is fixed, the optimal $\boldsymbol{v}_{b,k}$ and ${\boldsymbol{\mu }_{b,k}}$ can be explicitly determined by setting $\partial f/\partial{\boldsymbol{v}_{b,k}}$ and $\partial f/\partial{\boldsymbol{\mu }_{b,k}}$ to zero \cite{ref9}.

When optimizing $\mathbf{\Theta }_{b}$ with the obtained $\boldsymbol{v}_{b,k}$ and ${\boldsymbol{\mu }_{b,k}}$, we define $\boldsymbol{V}_b=\operatorname{diag}(\boldsymbol{\Theta}_b){\operatorname{diag}(\boldsymbol{\Theta}_b)^{\text{H}}}$ to simplify problem (\ref{Formulas4}), where $\boldsymbol{V}_b\succeq 0$ is a rank-one matrix. Based on this, problem (\ref{Formulas4}) can be reformulated as
\begin{equation}
	\begin{aligned}
		\label{Formulas5}
		\max_{\boldsymbol{V}_b}\; & f(\boldsymbol{V}_b)=\frac{1}{BK}\sum\nolimits_{b=1}^{B}\sum\nolimits_{k=1}^{K}\,c_6+c_7\sqrt{\text{Tr}(\boldsymbol{A}_1\boldsymbol{V}_b)}\\
		&\quad\quad\quad-\boldsymbol{\mu}_{b,k}^2(c_4^2\text{Tr}(\boldsymbol{A}_1\boldsymbol{V}_b)+c_5^2\text{Tr}(\boldsymbol{A}_2\boldsymbol{V}_b)+\sigma_{2}^{2})\\
		\text{s.t.} \;&\quad \text{C4}:\sum\nolimits_{b=1}^{B}\sum\nolimits_{k=1}^{K}{\eta}{c_{1}^2}{\boldsymbol{r}^{\text{T}}}{\boldsymbol{l}_{SR}}\text{Tr}(\boldsymbol{A}_3\boldsymbol{V}_b)+\\
		&\quad\quad\quad\quad\quad\quad{\eta}\sigma _{1}^{2}\text{Tr}(\boldsymbol{V}_b)+BMN{{P}_{C}}\le{E}_{max},\\
		&\quad\text{C5}:\boldsymbol{V}_{b_{i,i}}\le{({a}_{max}})^2 ,i\in [1\!:\!NM],
	\end{aligned}
\end{equation}
where ${c_6=\log_2(1+\boldsymbol{v}_{b,k})-\boldsymbol{v}_{b,k}}$, ${c_7=2{\boldsymbol{\mu}_{b,k}}\sqrt{1+{\boldsymbol{v}_{b,k}}}c_4}$, $\boldsymbol{A}_1=\operatorname{diag}(\boldsymbol{H}_{ru}^{\text{H}})\boldsymbol{H}_{sr}\boldsymbol{w}_{b,k}\boldsymbol{w}_{b,k}^{\text{H}}\boldsymbol{H}_{sr}^{\text{H}}\operatorname{diag}(\boldsymbol{H}_{ru})$, $\boldsymbol{A}_2=\boldsymbol{H}_{ru}\boldsymbol{H}_{ru}^{\text{H}}$, and $\boldsymbol{A}_{3}=\boldsymbol{H}_{sr}\boldsymbol{w}_{b,k}\boldsymbol{w}_{b,k}^{\text{H}}\boldsymbol{H}_{sr}^{\text{H}}$. To tackle the non-convexity introduced by the subtraction objective function, the SCA method\cite{ref11} is employed to exploit its lower bound. Accordingly, the first-order Taylor approximation is 
\begin{equation}
	\begin{aligned}
		\label{Formulas6}
		&\widetilde{f}(\boldsymbol{V}_b)={f}(\boldsymbol{V}_b)+\\
		& \text{Tr}((\frac{c_7\boldsymbol{A}_1}{2\sqrt{\boldsymbol{A}_1\boldsymbol{V}_b^{(q)}}}-\boldsymbol\mu _{b,k}^{2}(c_4^2\boldsymbol{A}_1+c_5^2\boldsymbol{A}_2))(\boldsymbol{V}_b-\boldsymbol{V}_b^{(q)})),
	\end{aligned}
\end{equation}
where $\boldsymbol{V}_b^{(q)}$ is the feasible point in the $q$-th iteration. To address the existence of the rank-one constraint, the penalty method \cite{ref12} is adopted here. Note that the rank-one constraint on $\boldsymbol{V}_b$ can be equivalently expressed as $\|\boldsymbol{V}_b\|_*-\|\boldsymbol{V}_b\|_2=0$, where $\|\boldsymbol{V}_b\|_*$ and $\|\boldsymbol{V}_b\|_2$ denote the nuclear norm and spectral norm of $\boldsymbol{V}_b$, respectively. Similarly, the SCA method is also adopted to address the non-convexity introduced by the subtractive form, leading to problem (\ref{Formulas7}) as follows:
\begin{equation}
	\begin{aligned}
		\label{Formulas7}
		\max_{\boldsymbol{V}_b} \quad &\widetilde{f}(\boldsymbol{V}_b)-\rho(\|\boldsymbol{V}_b\|_*-\|\boldsymbol{V}_b^{(q)}\|_2+\\
		&\text{Tr}\Bigl[u(\boldsymbol{V}_b^{(q)})u(\boldsymbol{V}_b^{(q)})^{\text{H}}(\boldsymbol{V}_b - \boldsymbol{V}_b^{(q)})\Bigr])\\
		\text{s.t.} \quad& \text{C4},\text{C5},
	\end{aligned}
\end{equation}
where $u(\boldsymbol{V}_b^{(q)})$ denotes the eigenvector corresponding to the largest eigenvalue of $\boldsymbol{V}_b^{(q)}$. $\rho>0$ represents the penalty factor, which is initially set to be a small value and updated iteratively by $\rho^{(q)}=\varepsilon\rho^{(q-1)}$, where $\varepsilon>1$ is a step size. A near-optimal solution for active RIS beamforming can be derived by solving (\ref{Formulas7}) iteratively with the updated penalty factor.
\subsubsection{Optimization of active RIS Orientation $\boldsymbol{r}$}
 we optimize $\boldsymbol{r}$ with the previously obtained ${\{\boldsymbol{w}_{b,k}\}_{b\in\mathbf{B}, k\in\mathbf{K}}}$ and $\{\mathbf{\Theta}_b\}_{b\in\mathbf{B}}$. Similar to Section~\ref{sec:III-B2}, the FP framework is employed to solve this subproblem in the form of multiple-ratio, yielding:
\begin{equation}
	\begin{aligned}
		\label{Formulas8}
		\max_{\boldsymbol{r},{\boldsymbol{v}_{b,k}},{\boldsymbol{\mu}_{b,k}}} &f(\boldsymbol{r},{\boldsymbol{v}_{b,k}},{\boldsymbol{\mu }_{b,k}})=\sum\nolimits_{b=1}^{B}\sum\nolimits_{k=1}^{K}{{{\log }_{2}}(1+{\boldsymbol{v}_{b,k}})}\\
		&-{\boldsymbol{v}_{b,k}}+2{\boldsymbol{\mu }_{b,k}}\sqrt{1+{\boldsymbol{v}_{b,k}}}{{c}_{8}}\sqrt{{\boldsymbol{r}^{\text{T}}}{\boldsymbol{l}_{sr}}{\boldsymbol{r}^{\text{T}}}{\boldsymbol{l}_{ru}}}\\
		&-\boldsymbol\mu _{b,k}^{2}({{c}_{8}^2}{\boldsymbol{r}^{\text{T}}}{\boldsymbol{l}_{sr}}{\boldsymbol{r}^{\text{T}}}{\boldsymbol{l}_{ru}}+c_{9}^2{\boldsymbol{r}^{\text{T}}}{\boldsymbol{l}_{ru}}+\sigma _{2}^{2})\\
		\text{s.t.}\quad& \text{C1},\text{C2},\text{C4},
	\end{aligned}
\end{equation}
where ${{c}_{8}=c_1{\boldsymbol{H}_{sr}^{\text{H}}}{\boldsymbol{\Theta}_{b}}\boldsymbol{H}_{sr}\boldsymbol{w}_{b,k}}$ and ${c_{9}=c_2\left\|{{\boldsymbol{H}_{ru}^{\text{H}}\boldsymbol{\Theta }}_{b}}\right\|\sigma_{1}}$. Similarly, $\boldsymbol{v}_{b,k}$ and $\boldsymbol{\mu}_{b,k}$ can be obtained by setting their derivatives to zero. Then, we let ${\boldsymbol{R}=[\boldsymbol{r};1][\boldsymbol{r};1]^\text{T}}$, which satisfies $\boldsymbol{R}\succeq 0$ and rank${(\boldsymbol{R})=1}$. Problem (\ref{Formulas8}) can be reformulated as
\begin{equation}
	\begin{aligned}
		\label{Formulas9}
		\max_{\boldsymbol{R}} \quad & f(\boldsymbol{R})=\frac{1}{BK}\sum\nolimits_{b=1}^{B}\sum\nolimits_{k=1}^{K}c_6+c_{10}\sqrt{\text{Tr}(\boldsymbol{A}_4\boldsymbol{R})}\\
		&\quad\quad\quad-\boldsymbol{\mu}_{b,k}^2(c_8^2\text{Tr}(\boldsymbol{A}_4\boldsymbol{R})+c_{9}^2\text{Tr}(\boldsymbol{A}_5\boldsymbol{R})+\sigma_{2}^{2})\\
		\text{s.t.} \quad &\text{C1:}\; \text{Tr}(\boldsymbol{R})=2,\\
		& \text{C2:} \; 0\le \text{Tr}(\boldsymbol{A}_5\boldsymbol{R}) \le 1,0\le \text{Tr}(\boldsymbol{A}_6\boldsymbol{R}) \le 1,\\
		& \text{C4:}\;\sum\nolimits_{b=1}^{B}\sum\nolimits_{k=1}^{K}{\eta}\left\|{c_{1}}{{\boldsymbol{\Theta }}_{b}\boldsymbol{h}_{sr}\boldsymbol{w}_{b,k}}\right\|^{2}\text{Tr}(\boldsymbol{A}_6\boldsymbol{R})\\
		&\quad\quad\quad\quad+{\eta}\left\|{{\boldsymbol{\Theta}}_{b}}\mathbf{I}\right\|^{2}\sigma _{1}^{2}+BMN{{P}_{C}}\le {E},\\
		&\text{C6:}; \boldsymbol{R}_{4,4}=1,
	\end{aligned}
\end{equation}
where\\
 $c_{10}=2{\boldsymbol{\mu}_{b,k}}\sqrt{1+\boldsymbol{\mu}_{b,k}}c_8$, $\boldsymbol{A}_4=\left[ \begin{matrix}{(\boldsymbol{l}_{sr}}\boldsymbol{l}_{ru}^{\text{T}}+\boldsymbol{l}_{ru}\boldsymbol{l}_{sr}^{\text{T}})/2&0&\\0&0\\\end{matrix} \right]$,  $\boldsymbol{A}_5=\left[ \begin{matrix}0&{\boldsymbol{l}_{ru}}/2  \\\boldsymbol{l}_{ru}^{\text{T}}/2&0\\\end{matrix} \right]$ and 
  $\boldsymbol{A}_6=\left[ \begin{matrix}0&{\boldsymbol{l}_{sr}}/2  \\\boldsymbol{l}_{sr}^{\text{T}}/2 & 0 \\\end{matrix} \right]$. Similar to (\ref{Formulas7}), the penalty function and SCA method are applied to resolve problem (\ref{Formulas9}) iteratively, which is given by
\begin{equation}
	\begin{aligned}
		\label{Formulas10}
		\max_{\boldsymbol{R}}\;& f(\boldsymbol{R})+\text{Tr}((\frac{c_{10}\boldsymbol{A}_4}{2\sqrt{\boldsymbol{A}_4\boldsymbol{R}^{(q)}}}-\boldsymbol{\mu}_{b,k}^2(c_8^2\boldsymbol{A}_4+c_{9}^2\boldsymbol{A}_5))\\
		&\quad\quad\quad\quad\quad (\boldsymbol{R}-\boldsymbol{R}^{(q)}))-\rho(\|\boldsymbol{R}\|_*-\|\boldsymbol{R}^{(q)}\|_2+\\ 
		&\quad\quad\quad\quad\quad \text{Tr}\Bigl[u(\boldsymbol{R}^{(q)})u(\boldsymbol{R}^{(q)})^\text{H}(\boldsymbol{R} - \boldsymbol{R}^{(q)})\Bigr])\\
		\text{s.t.}\;\;&\text{C1},\text{C2},\text{C4},\text{C6}.
	\end{aligned}
\end{equation}
\subsection{Outer-layer Optimization and Complexity Analysis}
In the outer-layer, we employ an iterative optimization framework to determine the optimal holding interval duration $K$. The process starts by initializing $K$ to its maximum value (i.e., the total communication period duration $T$) and progressively reducing its value. For each candidate $K$, the inner-layer optimization is implemented. The iteration of the outer-layer optimization continues until the solution corresponding to a specific $K$ violates constraint C4. Among all feasible candidates, the value of $K$ that yields the maximum objective value is selected as the optimal solution. The proposed algorithm for solving problem (\ref{Formulas6}) is described in Algorithm~\ref{algorithm1}. 

We further analyze the computational complexity of Algorithm 1. For the transmit beamforming, the computational complexity incurred by eigenvalue decomposition is $\mathcal{O}(L^3)$. According to \cite{ref9} and \cite{ref12}, the computational complexities of solving problems (\ref{Formulas7}) and (\ref{Formulas10}) are both given by $\mathcal{O}(I_{f}(I_{S}(MN)^{4.5}\log(1/\varepsilon)+2BK))$. Consequently, the overall computational complexity of Algorithm~\ref{algorithm1} is $\mathcal{O}(I_{K}I_{a}(2I_{f}(I_{S}(MN)^{4.5}\log(1/\varepsilon)+2BK)+BKL^3)$, where $I_{K}$ is the number of iterations that finding the optimal holding interval duration needs to run, $I_{a}$, $I_{f}$ and $I_{s}$ are the number of iterations by applying the AO algorithm, the FP framework and the SCA method, respectively.
\begin{algorithm}[!ht]
	\caption{Algorithm for solving Problem (\ref{Formulas6})}
	\label{algorithm1}
	\begin{algorithmic}[1] 
		\STATE Initialization: ${\{\boldsymbol{w}^{(0)}_{b,k}\}_{b\in\mathcal{B}, k\in\mathcal{K}}}$, $\{\mathbf{\Theta}^{(0)}_b\}_{b\in\mathbf{B}}$, $\boldsymbol{r}^{(0)}$, $q$, $\rho$, $\varepsilon$, ${K}$, convergence tolerance $\epsilon$, candidate sets $\mathcal{P}$ and $\mathcal{R}$.
		\STATE \textbf{repeat}
		\STATE \quad $K=K-1$.
		\STATE \quad\textbf{repeat}
		\STATE \quad\quad $q=q+1$.
		\STATE \quad\quad Update ${\{\boldsymbol{w}^{(q)}_{b,k}\}_{b\in\mathcal{B}, k\in\mathcal{K}}}$ by solving (\ref{Formulas3}).
		\STATE \quad\quad Update $\{\mathbf{\Theta}^{(q)}_b\}_{b\in\mathcal{B}}$ by solving (\ref{Formulas7}).
		\STATE \quad\quad Update $\boldsymbol{r}^{(q)}$ by solving (\ref{Formulas10}).
		\STATE \quad\textbf{until} the decrease of the objective function value in (\ref{Formulas2}) \hspace*{1em}is below $\epsilon$.
		\STATE \quad $\mathcal{P}=\mathcal{P}\cup\{{K,\{\boldsymbol{w}^{(q)}_{b,k}\}_{b\in\mathcal{B}, k\in\mathcal{K}}},\{\mathbf{\Theta}^{(q)}_b\}_{b\in\mathcal{B}},\boldsymbol{r}^{(q)}\}$ and \hspace*{1em}$\mathcal{R}=\mathcal{R}\cup\{\overline{R}^{(K)}\}$.
		\STATE \textbf{until} the power constraint C4 becomes to be infeasible.
		\STATE  Find the optimal variables achieving ${\overline{R}^*=\text{max}(\mathcal{R})}$ from $\mathcal{P}$.
	\end{algorithmic}
\end{algorithm}
\section{SIMULATION RESULTS}

In this section, we provide simulation results to demonstrate the performance of the proposed scheme. We assume that the coordinates of the user and obstacle are $\boldsymbol{p}_{U}=[0\mathrm{m},0\mathrm{m},0\mathrm{m}]$ and $\boldsymbol{p}_{O}=[0\mathrm{m},-100\mathrm{m},250\mathrm{m}]$, respectively. According to \cite{ref3}, we consider a LEO satellite at an altitude of $550$km that moves along a circular Keplerian orbit at a velocity of $v=\sqrt{\frac{\mu }{{{r}_{o}}}}$, where ${r}_{o}=7871\mathrm{km}$ and $\mu =3.986004\times {{10}^{5}}\mathrm{{km}^{3}/{{s}^{2}}}$ denote the orbital radius and Kepler constant, respectively. The total communication period is set as $T=100 \text{s}$, which corresponds to the time it takes for the satellite to move from the initial occlusion (i.e., the onset of communication link blockage) to the minimum communication elevation angle. The coordinate of the active RIS is set as $\boldsymbol{p}_{R}=[0\mathrm{m},80\mathrm{m},65.5\mathrm{m}]$, ensuring a VLoS link between the active RIS and the satellite. The communication frequency is $2\mathrm{GHz}$. We set ${{P}_{S}}=15\mathrm{dBW}$, ${{G}_{S}}=24.5\mathrm{dB}$, ${{G}_{U}}=10\mathrm{dB}$, ${{\sigma }_{1}}=-110\mathrm{dBW}$, ${{\sigma }_{2}}=-129\mathrm{dBW}$, ${{P}_{\text{C}}}=-10\mathrm{dBm}$, $\eta=1.25$, $a_{max}=10\mathrm{dB}$, $\ln(\xi_\mathrm{dB})\sim\mathcal{CN}(-0.6,0.4)$, $\kappa=3$, and $L=4$.

\begin{figure}[!t]
	\vspace{-2mm} 
	\centering
	\includegraphics[width=2.5in]{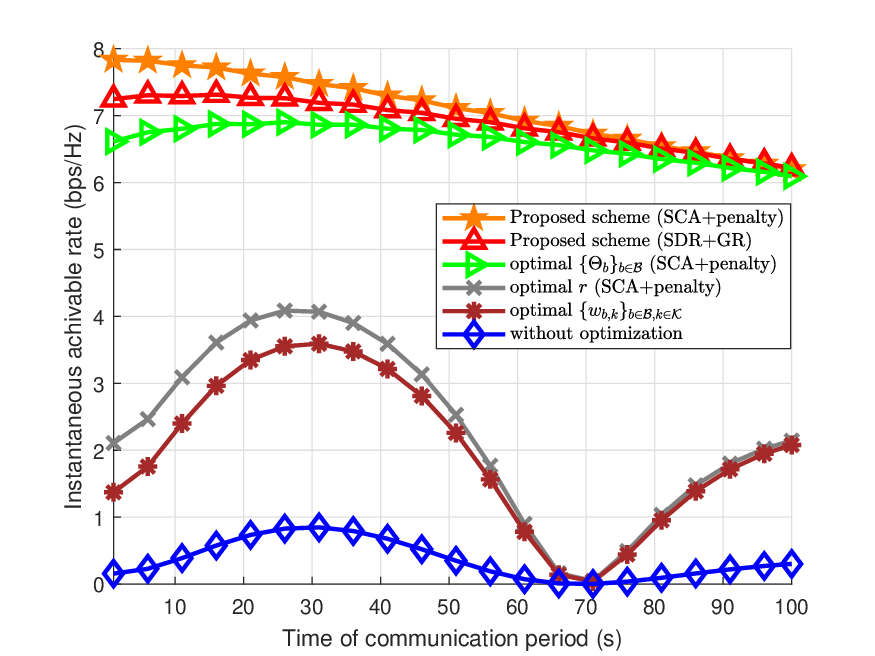}
	\caption{Instantaneous achievable rate with different optimization method.}
	\label{fig_3}
	\vspace{-3mm} 
\end{figure}

Fig. \ref{fig_3} illustrates the instantaneous achievable rate in each unit time slot during the communication period with $M=N=6$ and $K=10$. It is shown that the proposed scheme with SCA and penalty methods can achieve higher instantaneous achievable rates compared to the proposed scheme with the semidefinite relaxation (SDR) integrated with the Gaussian randomization (GR) approach. This superiority arises because the SCA scheme iteratively approaches the rank-one constraint, whereas the SDR scheme with the randomness of GR fails to tightly approximate the optimal solution. Moreover, both the partially optimized parameter scheme and the unoptimized scheme exhibit significant performance degradation compared to the proposed scheme, which validate the effectiveness of joint optimization of all variables.

\begin{figure}[!t]
	\vspace{-2mm} 
	\centering
	\includegraphics[width=2.5in]{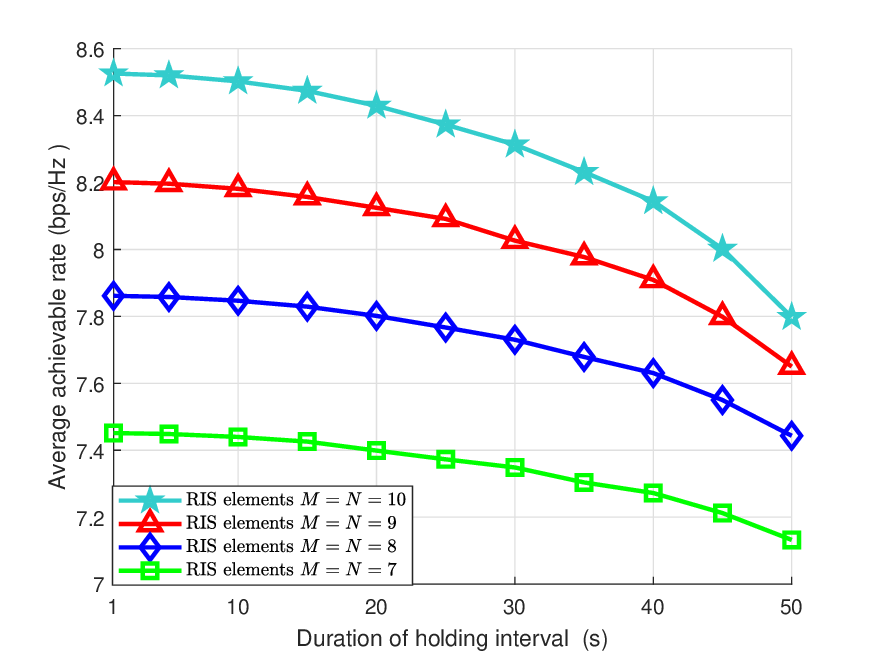}
	\caption{Average achievable rate versus the holding interval duration.}
	\label{fig_4}
	\vspace{-3mm} 
\end{figure}

Fig. \ref{fig_4} illustrates the average achievable rate during the communication period versus the holding interval duration. It is observed that the average achievable rate decreases as the holding interval duration increases, owing to the limited adaptiveness of long-term fixed RIS beamforming to fast channel variations. It is also shown that increasing the number of active RIS elements exacerbates the performance drop under long holding intervals. This is because larger active RIS arrays result in narrower beamwidths, making long-term fixed beamforming less adaptive to the large holding interval.

\begin{figure}[!t]
	\vspace{-2mm} 
	\centering
	\includegraphics[width=2.5in]{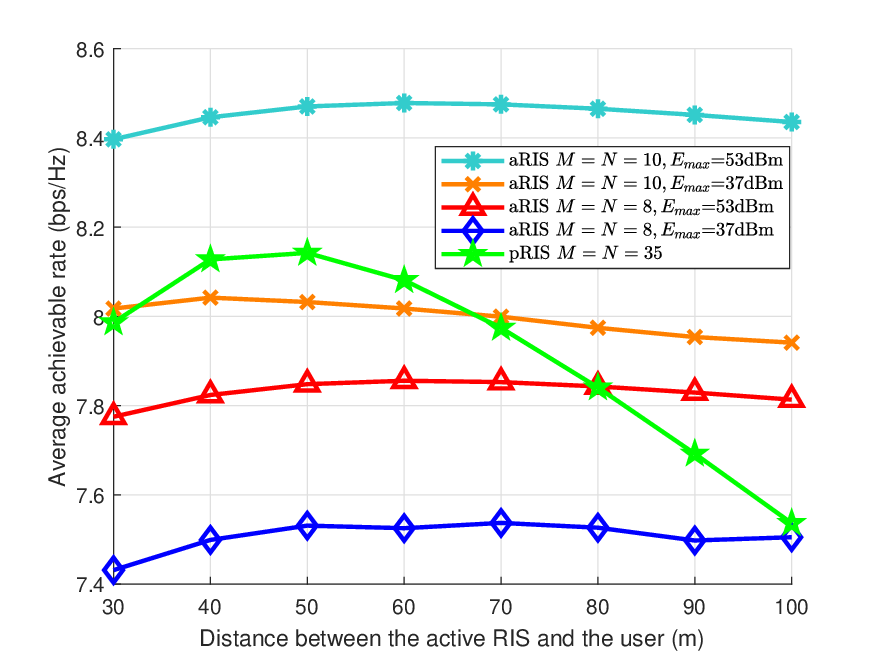}
	\caption{Average achievable rate versus the aRIS-User distance.}
	\label{fig_5}
	\vspace{-3mm} 
\end{figure}

Finally, we analyze the impact of the distance between the active RIS and the user on the average achievable rate under various numbers of elements and sizes of energy budgets, as illustrated in Fig. \ref{fig_5}. Specifically, compared to the passive RIS (pRIS), the proposed scheme with the active RIS can effectively reduce the number of required elements and enhance the average achievable rate by consuming additional power. It is also found that the active RIS exhibits lower sensitivity to the distance, implying its potential to support reliable communication over a broader coverage area. Furthermore, increasing the number of elements and enlarging the energy budget of the active RIS can further enhance system performance.


\section{CONCLUDING}
In this letter, we proposed the use of an active RIS to overcome communication interruptions caused by visibility obstructions in LEO satellite communication systems. To this end, we proposed a three-timescale optimization framework that jointly optimizes the transmit beamforming, the beamforming and direction vector of the active RIS. Furthermore, a two-layer solution algorithm was introduced to maximize the system’s achievable rate. Finally, numerical results have validated the effectiveness of the proposed scheme, demonstrating its superior performance in improving communication reliability and efficiency.

\bibliographystyle{unsrt}
\bibliography{myref}

\begin{thebibliography}{}
\bibliographystyle{IEEEtran}
\bibitem{ref1}
Z. Zheng et al., 
\newblock{“RIS-Enhanced LEO Satellite Communication: Joint Passive Beamforming and Orientation Optimization,”}
\newblock{\em{2022 IEEE Globecom Workshops}}., Rio de Janeiro, Brazil, 2022, pp. 874-879.

\bibitem{ref2}
W. Tang et al, \newblock{“Wireless Communications With Reconfigurable Intelligent Surface: Path Loss Modeling and Experimental Measurement,”} \newblock{\em{IEEE Trans. Wireless Commun}}., vol. 20, no. 1, pp. 421-439, Jan. 2021.

\bibitem{ref3}
B. Matthiesen et al.,
\newblock{“Intelligent Reflecting Surface Operation Under Predictable Receiver Mobility: A Continuous Time Propagation Model,”} \newblock{\em{IEEE Wireless Commun. Lett}}., vol. 10, no. 2, pp. 216-220, Feb. 2021.

\bibitem{ref4}
X. Tian, N. Gonzalez-Prelcic, and T. Shimizu, \newblock{“Enabling NLoS LEO Satellite Communications with Reconfigurable Intelligent Surfaces,”}
\newblock{\em{arXiv:2205.15528}}, 2022.

\bibitem{ref5}
X. Zhang et al,\newblock{“RIS-Aided MIMO Downlink Transmission for Ultradense LEO Satellite-Terrestrial Networks,”} \newblock{\em{IEEE Internet Things J}}., vol. 12, no. 11, pp. 15304-15318, June, 2025.

\bibitem{ref6}
R. Long, Y. -C. Liang, Y. Pei, and E. G. Larsson, 
\newblock{“Active Reconfigurable Intelligent Surface-Aided Wireless Communications,”}
\newblock{\em{IEEE Transactions on Wireless Communications}}., vol. 20, no. 8, pp. 4962-4975, Aug. 2021.

\bibitem{ref7}
Ge Y and Fan J. \newblock{“Active reconfigurable intelligent surface assisted secure and robust cooperative beamforming for cognitive satellite-terrestrial networks,”}
\newblock{\em{IEEE Trans. Veh. Technol}}., vol. 72, no. 3, pp. 4108-4113, March 2023.

\bibitem{ref8}
M. -M. Zhao et al.,
\newblock {“Intelligent reflecting surface enhanced wireless networks: Two-timescale beamforming optimization,”} 
\newblock{\em{IEEE Trans. Wireless Commun}}, vol. 20, no. 1, pp. 2-17, Jan. 2021.

\bibitem{ref9}
K. Shen and W. Yu, 
\newblock{“Fractional Programming for Communication Systems—Part II: Uplink Scheduling via Matching,”} 
\newblock{\em{IEEE Trans. Signal Process}}., vol. 66, no. 10, pp. 2631-2644, May, 2018.

\bibitem{ref10}
Q. Wu and R. Zhang,
\newblock{“Intelligent Reflecting Surface Enhanced Wireless Network via Joint Active and Passive Beamforming,”}
\newblock{\em{IEEE Trans. Wireless Commun}}., vol. 18, no. 11, pp. 5394-5409, Nov. 2019.

\bibitem{ref11}
V. Kumar et al. \newblock{“A Novel SCA-Based Method for Beamforming Optimization in IRS/RIS-Assisted MU-MISO Downlink,”} 
\newblock{\em{IEEE Wireless Commun. Lett.}}, vol. 12, no. 2, pp. 297-301, Feb. 2023.

\bibitem{ref12}
B. Lyu at al., 
\newblock{“Energy-Efficiency Maximization for STAR-RIS Enabled Cell-Free Symbiotic Radio Communications,”}
\newblock{\em{IEEE Trans. Cognit. Commun. Networking.}}, vol. 10, no. 6, pp. 2209-2223, Dec. 2024.


\bibitem{ref13}
H. E. Hammouti et al. {“Energy Efficient Aerial RIS: Phase Shift Optimization and Trajectory Design,”}
\newblock{\em{2024 IEEE Veh. Technol. Conf.}}, Singapore, Singapore, 2024, pp. 1-7.

\bibitem{ref14}
Lin Z et al., 
\newblock{“Refracting RIS-aided hybrid satellite-terrestrial relay networks: Joint beamforming design and optimization,”} 
\newblock{\em{IEEE Trans. Aerosp. Electron. Syst.}},
vol. 58, no. 4, pp. 3717-3724, Aug. 2022


\bibitem{ref15}
X.-D. Zhang,\newblock{\em{Matrix Analysis and Application.}}, Cambridge, U.K.: Cambridge Univ. Press, 2017.
\end{thebibliography}

\end{document}